\documentclass[review]{elsarticle}

\usepackage{lineno,hyperref}
\usepackage{multirow}
\usepackage{longtable}
\usepackage{graphicx}

\usepackage{algorithm}
\usepackage{algpseudocode}
\usepackage{float}
\usepackage{tikz}
\usepackage{amssymb}

\usepackage{natbib} 
\usepackage{mathtools}
\usepackage[all,cmtip]{xy}

\usepackage{subcaption}

\usepackage{amsfonts,amssymb,amsmath,amsthm,enumerate,mathrsfs,tikz,xparse}

\journal{European Journal of Operational Research}

\biboptions{authoryear}

\newtheorem{thm}{Theorem}

\newtheorem{cor}{Corollary}
\newtheorem{obs}{Observation}

\long\def\comment#1\endcomment{}

\begin{document}

\begin{frontmatter}

\title{Exact and Heuristic Algorithms for the Domination Problem} 

\author{Ernesto Parra Inza\fnref{myfootnote1}}
\ead{eparrainza@gmail.com}
\fntext[myfootnote1]{ Centro de Investigación en Ciencias, UAEMor, Cuernavaca, Morelos, México.}

\author[myfootnote1]{Nodari Vakhania\corref{mycorrespondingauthor}}
\cortext[mycorrespondingauthor]{Corresponding author}
\ead{nodari@uaem.mx}

\author{José María Sigarreta Almira\fnref{myfootnote2}}
\ead{josemariasigarretaalmira@hotmail.com}
\fntext[myfootnote2]{Facultad de Matemáticas, UAGro, Acapulco de Juárez, Guerrero, México.}

\author{Frank Angel Hernández Mira\fnref{myfootnote3}}
\ead{fmira8906@gmail.com}
\fntext[myfootnote3]{Centro de Ciencias de Desarrollo Regional, UAGro, Acapulco de Juárez, Guerrero, México.}


\begin{abstract}
In a simple connected graph $G=(V,E)$, a subset of vertices $S \subseteq V$ is a 
dominating set if any vertex $v \in V\setminus S$ is adjacent to some vertex $x$ 
from this subset. A number of real-life problems can be modeled using this problem
which is known to be among the difficult NP-hard problems in its class. We formulate 
the problem as an integer liner program (ILP) and compare the performance with the two earlier existing exact state-of-the-art algorithms and exact implicit enumeration and heuristic algorithms that we propose here.
Our exact algorithm was able to find optimal solutions 
much faster than  ILP  and the above two exact algorithms for middle-dense instances.  
For graphs with a considerable size, our heuristic algorithm
was much faster than both, ILP and our exact algorithm. It found an optimal 
solution for more than half of the tested instances, whereas it improved the earlier 
known state-of-the-art solutions for almost all the tested benchmark instances. Among the 
instances where the optimum was not found, it gave an average approximation error of $1.18$. 
\end{abstract}


\end{frontmatter}


\section{Introduction}

Finding a minimum dominating set in a graph is a traditional discrete optimization problem. In a simple connected graph $G=(V,E)$ with $|V|=n$ vertices and $|E|=m$ edges, a subset of vertices $S \subseteq V$ is a \textit{dominating set} in graph $G$ if any vertex $v \in V$ is \textit{adjacent} to some vertex $x$ from this subset (i.e., there is an edge $(v,x)\in E)$ unless vertex $v$ itself belongs to set $S$. Any subset $S$ with this property will be referred to as a \textit{feasible solution}, whereas any subset of vertices from set $V$ will be referred to as a \textit{solution}.  The number of vertices in a solution will be referred to as its \textit{size} (or \textit{order}). The general objective is to find an \textit{optimal solution}, a feasible solution with the minimum possible size $\gamma(G)$.\\

The domination problem is known to be \textit{NP-hard} \cite{Garey}, and is among the hardest problems in the family. Not only it is not fixed parameter tractable but it does not admit fixed-parameter tractable approximation algorithms\cite{Chalermsook2020,feldmann2020survey,lin2019simple}.
A complete enumeration of all feasible solutions would imply an exponential cost of $O(2^{n})$ (for a graph with only $20$ nodes, 
such an enumeration would take centuries on modern computers). 
To the best of our knowledge, the only exact  algorithms which do better than a complete enumeration  were  described in \cite{van2011exact} and \cite{iwata2012faster}. The authors show that their 
algorithms run in times $O(1,4969^{n})$ (\cite{van2011exact} ) and $O(1.4689^{n})$ (\cite{iwata2012faster}), but they did not report any experimental study. The latter two bounds are obviously better than $2^{n}$, but they remain 
impractical (for instance, for graphs with 100 vertices, modern computers will require almost 7 years to enumerate all feasible solutions). In this work, we will give a 
detailed analysis of the practical behavior of these algorithms.\\
 
As to heuristic solution methods, 
\cite{campan2015fast} and \cite{eubank2004structural} present some heuristics 
with some experimental study.  The authors do not present approximation errors
given by their heuristic algorithms, but they considered large-scale 
social network instances with up to 100000 vertices. For the weighted set cover
problem, \cite{chvatal1979greedy} proposed an approximation algorithm with 
approximation ratio $\ln\left(\frac{n}{OPT}\right) + 1 + 
\frac{1}{OPT}$. \cite{parekh} adopted the latter algorithm for the domination problem,  
and  showed that the cardinality of the dominating set created by that algorithm is 
upper bounded by $n+1-\sqrt{2m+1}$. Recently, an improved two-stage heuristic algorithm was proposed in \cite{Mira2021}, that also includes a detailed overview of the state-of-the-art solutions methods and results on domination problems.\\  

Some generalizations of the domination problem have been considered. In the so-called $k$-dominating set, every vertex of $G$ must possess at least $k$  neighbors from this set or be in the set. In the {\em total} dominating set, every vertex must have at least one neighbor from that set. In both variations, the objective is to minimize the number of vertices. In the weighted version of the total domination problem, the edges and vertices are assigned positive weights and the
objective is  to find a total dominating set $S$ with the minimum total weight (defined as the sum of all vertex and edge weights in the sub-graph induced by $S$, 	plus the minimum weight of an edge joining a vertex $x$ not in $S$ with a vertex in $S$, for every vertex $x\not\in S$). In \cite{ALVAREZMIRANDA2021105157} an algorithm that optimally solved the weighted total domination domination problem for  instances with up to 125 vertices within a time-limit of 30 minutes was suggested. In \cite{CORCORAN2021105368}, the heuristic from \cite{parekh} was adopted for the $k$-domination problem and the obtained results were used to solve the facility location problems in street networks. \cite{haynes2017domination} and \cite{JOSHI1994352} apply dominating sets and $k$-dominating sets, respectively,  in the solution of the facility location problem. \cite{haynes2017domination} it is described how other related graph-theoretic 
problems, including set cover, maximum independent set and chromatic number, can be reduced to the dominance problem. \cite{liao2005power} apply the dominating sets to monitor electric power system to minimize the phase measurement units. \cite{haynes2002domination} also study domination in graphs applied to electric power networks. Different forms of graph domination have been used  to model clustering in wireless ad hoc networks \cite{balasundaram2006graph}, \cite{wan2004distributed}, \cite{wu2003power} and \cite{wu2002extended}. Another area of application for dominating sets is the robustness analysis of food webs \cite{RTParra}. \cite{PINACHODAVIDSON2018860} proposed integer linear programming models and two greedy heuristics for a weighted version of independent domination problem. The solutions delivered by the heuristics were compared to exact solutions obtained by the integer linear programs for instances with up to 100 nodes.\\

{\bf Our Contributions.} 
We propose an exact implicit enumeration and an approximation 
heuristic algorithms, and give approximation ratios for the heuristic. We 
also formulate the problem as an integer liner program (ILP) and test 
its performance vs our exact and heuristic algorithms, and vs 
the earlier mentioned exact algorithms from \cite{van2011exact} and 
\cite{iwata2012faster} which we have also coded.\\ 

We verified  the algorithms 
for more than 600 existing benchmark instances  \cite{bdparra1} and for
500  problem instances that we generated (they are publicly available at 
\cite{bdparra2}, together with the codes of the  algorithms from 
\cite{van2011exact} and \cite{iwata2012faster}, the codes of our two
algorithms and the computational results reported in this work). We present a 
detailed experimental analysis in Section 6. According to our experimental study, 
the algorithms \cite{van2011exact} and \cite{iwata2012faster} perform better for
quasi-trees with up to 120 vertices, whereas ILP was better for very dense graphs. 
For average middle-dense graphs, the algorithms from \cite{van2011exact} and \cite{iwata2012faster} were able to solve problem instances with up 
to 300 vertices within the range of 8 hours. For these instances, our exact 
algorithm, in average, was about 18 times faster (for every tested instance, we compared the solution created by our exact algorithm with the best, among the two solutions obtained by the latter two algorithms). For middle-dense graphs with more than 250 vertices, our exact algorithm was better than ILP, but ILP behaved better 
for smaller sized graphs. The former algorithm was able to find optimal solutions, 
in average, 874 times faster than  ILP. For graphs with a considerable size 
(from 600 to 2000 vertices), the heuristic was, in average about 186 
times faster than ILP, and about 49.8 times  faster than our exact algorithm. The heuristic found an optimal solution for 61.54\% of the instances, whereas, among the instances where the optimum was not found, the approximation error was $1.18$. It has improved the earlier known state-of-the-art solutions in 98.62\% of the analyzed instances.\\ 

We complete this section with the description of brief 
construction ideas of our 
algorithms. They rely on earlier known lower and upper bounds on the number 
of vertices in an optimal solution. As an initial lower bound $L$, we take the maximum
among lower bounds proposed in  \cite{haynes2017domination} and \cite{cabrera2020total}. 
We let the initial upper bound $U$ be the size of the feasible solution constructed 
by the heuristic from \cite{Mira2021}. Our exact algorithm  enumerates the solutions 
of the size from the range $[L,U]$ applying binary search in this interval. 
Both algorithms enumerate solutions of the
same size in a specially determined priority order (which is different for 
each algorithm).  The heuristic algorithm combines depth-first 
search with a breadth-first search. Special type of partial solutions of size 
$\beta = \lfloor\alpha (U-s)\rfloor + s$, for $\alpha \in \mathbb{R}$ with $  0 < \alpha < 1 $, are enumerated using breadth-first search, where $s$ is the total number of support vertices (vertices adjacent to at least one vertex of degree one) in graph $G$. The extensions of these solutions are generated in the depth-first fashion.\\

In the next section we give the necessary preliminaries. In Section 3 we describe 
an ILP formulation of the domination problem, and in Sections 4 and 5 we  describe  
our exact and heuristic algorithms, respectively. 
Section 6 presents the conducted  experimental study, and Section 7 
contains a few final remarks. Preliminary version of this work was presented at
the First On-Line Conference on Algorithms IOCA 2021 \cite{Nodari2021}.

\section{Basic properties}

In this section we define some basic properties on graphs that we use in our
implicit enumeration algorithm. We accompany them with the necessary definitions.\\

The \emph{diameter} $d(G)$ of graph $G$ is the maximum number of edges on the shortest path between any pair of vertices in that graph, and the \emph{radius} $r(G)=\min _{v \in V(G)} \max _{u \in V(G)- v} dis(v,u)$, where $dis(u,v)$ denotes the number of edges on the shortest path connecting $v$ and $u$. 
A {\em leaf} vertex is a degree one vertex, 
and a {\it support} vertex is a vertex adjacent to a leaf.  The sets of support and leaf vertices in graph $G$ will be denoted by  $Supp(G)$ and $Leaf(G)$, respectively, and we will let $s=|Supp(G)|$ and $l=|Leaf(G)|$. 
The degree of a vertex with the maximum number of neighbors in graph $G$ is
denoted by $\Delta(G)$.\\ 

The initial lower bound $L$ on the size of an optimal solution that our algorithms 
employs is obtained based on the following known results.

\begin{thm}\label{teorema1}
	\cite{haynes2017domination} If $G(V,E)$ is a connected graph of order $n$, 
	then $ \gamma(G)\geq\frac{n}{\Delta(G)+1}$.
\end{thm} 

\begin{thm}\label{teorema2}
	\cite{haynes2017domination} If $G(V,E)$ is a connected graph, 
	then $ \gamma(G)\geq \frac{2r(G)}{3}$ and $ \gamma(G)\geq \frac{d(G)+1}{3}$.
\end{thm}

\begin{thm}\label{teorema3}
	\cite{haynes2017domination} If $G(V,E)$ is a connected graph of order $n$, 
	then $s \leq \gamma(G) \leq n- l$.
\end{thm}

The next corollary is an immediate consequence of the Theorems \ref{teorema1}, \ref{teorema2} and \ref{teorema3}.

\begin{cor}\label{cororario1}
	If $G(V,E)$ is a connected graph of order $n$, then,  $$L=\max\left\lbrace \frac{n}{\Delta(G)+1},\frac{2r(G)}{3},\frac{d(G)+1}{3},s\right\rbrace $$ is a lower bound on the size of an optimal solution.
\end{cor}

From Theorem 3, we obtain the following corollary.

\begin{cor}\label{cororario11}
	In a connected graph $G$,  there exists at least one dominating set $S$ with cardinality $\gamma(G)$ such that $Supp(G) \subseteq S$ and $Leaf(G)\nsubseteq S$.
\end{cor}

\section{ILP formulation}

The ILP formulation of the domination problem, that we propose
here, is similar to ones from \cite{fan2012solving} and \cite{PINACHODAVIDSON2018860} for the connected dominating set problem and weighted independent domination problem, respectively. Given graph $G=(V,E)$ of order $n$, the neighborhood matrix of $G$, 
$A=(a_{ij})_{n\times n}$, is defined as follows: $ a_{ij} = a_{ji} = 1 $ if $(i,j) \in E(G)  $ or $ i=j $, and $ a_{ij} = a_{ji} = 0 $ otherwise. Given a dominating set of cardinality $\gamma(G)$ and $i \in V(G)$, we define our decision variables   as follows:

\begin{equation}
	x_{i}= \left\{  \begin{array}{lcl}
		1, &  \mbox{ if }  i \in S\\
		0, &  \mbox{ otherwise, }   
	\end{array}
	\right.
\end{equation}

The dominating set problem can now be formulated as follows:

\begin{equation} \label{fo}
	\min \sum_{i=1}^{n}x_{i}=\gamma(G),
\end{equation}

\begin{equation}\label{st}
subject \; to:\; \sum_{j=1}^{n}a_{ij}x_{j} \geq 1, \; x_{j}\in \lbrace 0,1\rbrace , \; \forall i \in V(G),
\end{equation}

Here the objective function is $ \sum_{i=1}^{n}x_{i}$, that we
wish to minimize (\ref{fo}). We have $n$ restrictions (\ref{st}), guarantying that  
for each $ j \in V (G) $, at least one of the vertices in $ S $ must be adjacent to $ j $ or $ j \in S$.

\section{The implicit enumeration algorithm}
Our exact implicit enumeration algorithm finds 
an optimal solution enumerating specially determined subsets of feasible
solutions. We obtain the initial upper bound $U$ on the size of a feasible 
solution by the approximation algorithm from \cite{Mira2021}.  
Given the lower bound $L$ from Corollary \ref{cororario1} and this upper bound, we restrict our search for feasible solutions with the size in the range $[L,U]$
using binary search.\\ 

The solutions of the size $\nu\in [L,U]$ are generated and tested for the feasibility based on the specially formed priority list of solutions. The sizes of feasible solutions are derived by binary division search accomplished in the interval $[L, U]$.
Typically, we choose $(L+3U)/4$ as the starting point in our search.
For each created solution $\sigma$ of size $\nu$, feasibility condition is verified, i.e., it is verified if the solution forms a dominating set. Below we describe the general framework of the algorithm. In the next subsection, we specify how the priority lists are created. \\

Let $\sigma$ be the current solution of size $\nu$ (initially, $\sigma$ is the
feasible solution delivered by the algorithm from \cite{Mira2021}): 

\begin{itemize}
	\item If solution $\sigma$ is feasible, then the current upper bound $U$ is updated to $\nu$.
	The algorithm proceeds with the next (smaller) trial value $\nu$   from the interval $[L,\nu)$ derived by binary search procedure. If all trial $\nu$s were already tested, then $\sigma$ is an optimal solution. The algorithm returns solution $\sigma$ and halts.
	
	\item If the current solution $\sigma$ of size $\nu$ is not feasible, then  $Procedure\_Next(\nu)$ is called and the next to $\sigma$ solution of size $\nu$ from the corresponding priority list is tested. 
	
	\item If $Procedure\_Next(\nu)$ returns $NIL$, {\it i.e.,}, all the solutions of size $\nu$ were already tested for the feasibility (none of them being feasible), the current lower bound $L$ is updated to $\nu$. The algorithm proceeds with the next (larger) trial value $\nu$  from the interval $[\nu, U)$ derived by  binary search procedure. If all trial $\nu$s were already tested, then $\sigma$ is an 	optimal solution. The algorithm returns solution $\sigma$ and halts. 
\end{itemize}

\subsection{$Procedure\_Next(\nu)$ and $Procedure\_Priority\_LIST()$}

Next, we describe in which order the solutions of a given size $\nu$ are considered
during the enumeration process. For each trial value $\nu \in [L, U]$, the solutions of size at most $\nu$ are generated in a special priority order that is intended to help in a faster convergence to a feasible solution. Two procedure are employed. Subroutine $Procedure\_Priority\_LIST()$ generates a priority list of vertices. The earlier included vertices in this list ``potentially cover'' a major number of yet uncovered vertices.  Based on the order determined
by $Procedure\_Priority\_LIST()$, $Procedure\_Next(\nu)$ determines the solution $\sigma_h(\nu)$
of the current trial size $\nu$ of iteration $h$. First we describe $Procedure\_Priority\_LIST()$.\\
  
Neither vertices from set $Supp(G)$ nor vertices from set $Leaf(G)$ are considered while forming the priority list of vertices. The vertices from the first set are to be included into any dominating set, hence they form part of any solution that is considered during the enumeration process; hence, none of the vertices from set $Leaf(G)$ need to be included into any formed solution (see Corollary \ref{cororario11}). \\ 

The priority list of vertices is formed based on their {\it active degrees}. The notion of an  active degree was introduced in \cite{Mira2021}. The active degree of a vertex is determined dynamically at every iteration $r$. Let $LIST_r$ be the priority list of iteration $r$ in $Procedure\_Priority\_LIST()$ (initially, $LIST_0:=\emptyset$). The active degree of vertex $v$ at iteration $r$ is the number of neighbors of vertex $v$ in set $V(G) \setminus \{LIST_{r-1} \cup Supp(G)\}$ not counting the neighbors of vertex $v$ which are adjacent to a vertex in $\{LIST_{r-1} \cup Supp(G)\}$.\\

Suppose $S$ is  a feasible solution (a dominating set). Priority list of vertices is iteratively formed by the vertices from the set $S \setminus (Supp(G)\cup Leaf(G))$ sorted in non-increasing order of their active degrees. The remaining vertices, i.e., ones from the set $V\setminus \{S \cup Supp(G)\cup Leaf(G)\}$, are iteratively inserted in non-increasing order of their degrees in graph $G$. A formal description of the procedure follows. 

\medskip
\begin{algorithm}[H]
	\fontsize{10pt}{.3cm}\selectfont
	\caption{Procedure\_Priority\_LIST}\label{alg_list}
	\begin{algorithmic}
		\State Input: A graph $G(V,E)$ and a feasible solution (dominating set) $S$.
		\State Output: Priority list $LIST := LIST_{r}$.
		
		\State $r := 0$; 
		\State $LIST_{r} := \emptyset$;
		\State $S_r:= S \setminus (Supp(G)\cup Leaf(G))$;  
		
		\{ iterative step \} 
		
		\While{$S_r \neq \emptyset$} 
		\State $r := r+1$; 
		\State $v_r$ := any vertex of set $S_{r-1}$ with the maximum active degree; 
		\State $S_r := S_{r-1} \setminus \{v_r\}$;
		\State $LIST_{r} := LIST_{r-1} \cup \{v_r\}$;
		\EndWhile
		
		\State $D_r:= V \setminus (S \cup Leaf(G))$; 
		\While{$D_r \neq \emptyset$} 
		\State $r := r+1$; 
		\State $v_r$ := any vertex of set $D_{r-1}$ with the maximum degree; 
		\State $D_r := D_{r-1} \setminus \{v_r\}$;
		\State $LIST_{r} := LIST_{r-1} \cup \{v_r\}$;
		\EndWhile
		
	\end{algorithmic}
	
\end{algorithm}
\medskip 

\bigskip

Now we describe $Procedure\_Next(\nu)$ that, iteratively, verifies if the currently formed solution $\sigma_h(\nu)$ is feasible. (Note that the feasibility of every generated solution of a given size can be verified in time $O(n)$.) If it is feasible, then it returns that solution and continues with the next trial $\nu$;  the procedure  halts if all trial sizes have been already considered. For every new trial value $\nu$, the first solution of that size contains all the vertices from set $Supp(G)$ complemented by the first $\nu-|Supp(G)|$ vertices from the list $LIST_r$ delivered by $Procedure\_Priority\_LIST()$.\\ 

If solution $\sigma_h(\nu)$ is not feasible, the next solution $\sigma_{h+1}(\nu)$ of size $\nu$ is obtained from solution $\sigma_h(\nu)$ by a vertex interchange as follows. Let $v \notin \sigma_h(\nu)$ be the next vertex from the priority list, and let $v' \in \sigma_h(\nu)$ be the vertex with the smallest active degree. Vertex $v'$ is substituted by vertex $v$  in solution
$\sigma_{h+1}(\nu)$ (note that the active degree of $v$ must be less than that of vertex $v'$):  
\begin{equation*}
	\sigma_{h+1}(\nu):=(\sigma_h(\nu)\setminus\{v'\}) \cup \{v\}. 
\end{equation*}

\bigskip
\medskip
\begin{algorithm}[H]
	\fontsize{10pt}{.3cm}\selectfont
	\caption{Procedure\_Next\_Aux}\label{alg_nex_aux}
	\begin{algorithmic}
		\State Input: $\nu^\prime$, $\sigma_h(\nu^\prime)$, $LIST$, index\_$LIST$, $Supp(G)$, $h$.
		\State Output: $\sigma^*(\nu)$ or $ NIL $.
		\If{$\nu^\prime = 0$}
		\If{$Supp(G) \cup \sigma_h(\nu^\prime)$ is feasible solution}
		\State $\sigma^*(\nu) := Supp(G) \cup \sigma_h(\nu^\prime)$;
		\State return $\sigma^*(\nu)$;
		\EndIf
		\State return $\sigma_{h-1}(\nu^\prime)$;
		\EndIf
		\State $i := $index\_$LIST$;
		\While{$i\leq |LIST|$} 
		\State Add $LIST[i]$ at the end of $\sigma_h(\nu^\prime)$; 
		
		
		\State $ \sigma_{h+1}(\nu^\prime) = Procedure\_Next\_Aux(\nu^\prime-1, \sigma_h(\nu^\prime), LIST, i+1, Supp(G), h+1) $
		
		\If{$Supp(G) \cup \sigma_h(\nu^\prime)$ is feasible solution}
		\State $\sigma^*(\nu) := Supp(G) \cup \sigma_h(\nu^\prime)$;
		\State return $\sigma^*(\nu)$;
		\Else
		\State Remove the last element in $\sigma_h(\nu^\prime)$;
		\State $i := i+1$;
		\EndIf		
		\EndWhile
		\State return $ NIL $;
	\end{algorithmic}
	
\end{algorithm}
\medskip 

\medskip
\begin{algorithm}[H]
	\fontsize{10pt}{.3cm}\selectfont
	\caption{Procedure\_Next($\nu$)}\label{alg_nex}
	\begin{algorithmic}
		\State Input: $\nu$, $LIST$, $Supp(G)$.
		\State Output: $\sigma^*(\nu)$ or $ NIL $.
		
		\State $h := 0$;
		\State index\_$LIST:=0$;
		\State $\nu^\prime:=\nu - Supp(G)$;
		\State $\sigma_h(\nu):=\emptyset$;
		\State return Procedure\_Next\_Aux($\nu^\prime$, $\sigma_h(\nu)$, $LIST$, index\_$LIST$, $Supp(G)$,  $h$);
		
	\end{algorithmic}
	
\end{algorithm}
\medskip 

Now we can give a formal description of our enumeration algorithm.
\medskip
\begin{algorithm}[H]
	\fontsize{10pt}{.3cm}\selectfont
	\caption{Algorithm\_BDS}\label{alg1}
	\begin{algorithmic}
		\State Input: A graph $G$.
		\State Output: A $\gamma(G)$-$set$ $S$.
		
		\State $Supp(G) :=$ Set of support vertex of graph $G$;
		\State $Leaf(G) :=$ Set of leaf vertex of graph $G$;
		\State $L := \max\{\frac{n}{\Delta(G)+1},\frac{2r}{3},\frac{d+1}{3},|Supp|\}$;
		\State $S:=\sigma$; \hspace{.3cm} \{Feasible solution proposed in \cite{Mira2021}\}
		\State $U := |S|$;
		\State $\nu := \lfloor(L+3U)/4\rfloor$;
		
		\State $LIST := Procedure\_Priority\_LIST(G,S)$;
		
		\{ iterative step \}

		\While{$U-L>1$} 
		
		\If{$Procedure\_Next(\nu)$ returns $NIL$}
		\State $L := \nu$;
		\State $\nu := \lfloor(L+3U)/4\rfloor$;
		\Else  \hspace{.3cm} \{ A feasible solution was found ($\sigma_h(\nu)$)\}
		\State $U := \nu$;
		\State $\nu := \lfloor(L+3U)/4\rfloor$;
		\State $LIST := Procedure\_Priority\_LIST()$; 
		\EndIf
		
		\EndWhile
		
	\end{algorithmic}
\end{algorithm} 
\bigskip

Let $s=|Supp(G)|$ and $l=|Leaf(G)|$. Below we give a brutal worst-case
 bound on the running time of our implicit enumeration algorithm.
	\begin{thm}\label{lemma1}
		A worst-case time complexity of Algorithm \ref{alg1} is
		\begin{equation*} 
			O\left(n\log(U-L)\binom{n-s-l}{\nu-s}\right).
		\end{equation*}	
	\end{thm}

\begin{proof}
Since binary search in the interval $[L,U]$ is carried out, the total number of external iterations in Algorithm \ref{alg1} (i.e., the number of different sizes $\nu$) is at most $\log(U-L)$. For a given size $\nu$, the number of the generated solutions of that size is bounded by $\binom{n-s-l}{\nu-s}$. Indeed, in the worst case, all solutions of cardinality $\nu$ are considered. By Corollary \ref{cororario11}, the set $Supp(G)$ forms part of all generated solutions, whereas no leaf vertex belongs to any created solution, and hence $\binom{n-s-l}{\nu-s}$ is an upper bound on the number of solutions of size $\nu$ that the algorithm creates. To establish the feasibility of solution $\sigma_h(\nu)$, $Procedure\_Next(\nu)$ verifies if every vertex $x \in V(G)$ is in $\sigma_h(\nu)$ or if it is adjacent to a vertex in $\sigma_h(\nu)$, which clearly takes time $O(n)$. Summing up the above, we have an overall bound $ O\left(n\log(U-L)\binom{n-s-l}{\nu-s}\right) $ on the cost of the algorithm.
\end{proof}

We may complement the above proposition with an intuitive analysis reflecting the worst-case behavior of our exact algorithm. For the purpose of this estimation, 
let us assume that $\nu = \lfloor(U+L)/2\rfloor$ (note that the maximum number of combinations is reached for this particular $\nu$). We may also express $\nu$ in terms of $n$ as $\nu= \lfloor(\frac{n}{2}+1)/2\rfloor=\frac{n}{4}$ using $U = n/2$, $L = 1$, $s = 0$ and $l = 0$ in case there is a feasible solution of size $\nu=n/4$, then $O\left(n\log(\dfrac{n}{2}-1)\binom{n}{n/4}\right)$. If there is no such a solution, the algorithm increases the current $ \nu $. Then in the next iteration, the algorithm will enumerate the possible solutions with $ \nu=n/3 $, $O\left(n\log(\dfrac{n}{2}-1)\binom{n}{n/3}\right)$. Similarly, there exists a feasible solution of size $n/3$ or not. If not,  the algorithm proceeds with the solutions of size $\frac{5}{12}n$, and so on. \\

The above analysis, reflects a brutal worst-case behavior of our algorithm, but  
does not convey its practical performance (see  Section 6 for details).

\section{The heuristic algorithm}

We need a few additional notations to describe our heuristic algorithm. 
Let $\sigma$ be the feasible solution obtained by the greedy algorithm from \cite{Mira2021}. We define an auxiliary parameter
\begin{equation} \label{beta}
	\beta = \lfloor\alpha (U-s)\rfloor + s ,
\end{equation} 
for $  0 < \alpha < 1 $ ($s=|Supp(G)|$). $\beta$ is the size of a {\em base solution} $ \sigma^h(\beta) $, a (partial) solution that serves us as a basis for the construction of some larger sized solutions. These solutions, which will be referred to as {\em extensions} of solution $ \sigma^h(\beta) $,  share the $ \beta $ vertices with solution $ \sigma^h(\beta) $.\\

\medskip

The set of vertices in any base solution contains
all support vertices from set $Supp(G)$. It is completed to a set of
$\beta$ vertices according to one of the following alternative rules:
\begin{enumerate}
	\item The first $  \beta - s $ non-support vertices from solution $ \sigma $. 
	\item  $\beta - s$ randomly generated non-support vertices from solution $ \sigma $. 
	\item  $\beta - s$ randomly generated vertices from set $V \setminus \{ Supp(G)\cup Leaf(G) \}$.\\
\end{enumerate}

For each of these options, the vertices in a newly determined base solution are selected in such a way that it does not coincide with any of the earlier formed base solutions. The procedure creates the first base solution by rule \textit{(1)} or rule \textit{(2)}. Next base solutions are obtained by rule \textit{(3)} unless the last such generated base solution coincides with an earlier created one. If this happens, then the remaining base solutions are created just in the lexicographic order.\\

Every base solution is iteratively extended by one vertex per iteration 
as long as the latest considered feasible solution is 
not feasible. Either  \textit{(i)} the base solution or 
some its extension  turns out to be feasible or \textit{(ii)} the extension 
of size $ U-1 $ is not feasible. In case \textit{(ii)}, the next base solution 
of size $ \beta $ is constructed and the newly created base solution is similarly extended. In case \textit{(i)}, the current upper bound $U$ and  the parameter $ \beta $ are
updated (see formula \ref{beta}) and the procedure similarly processes the first base 
solution of the new size $ \beta $.\\ 

The algorithm halts when it processes all base solutions
of the current cardinality $\beta$ without finding a feasible solution 
(i.e., none of the base solutions of size $\beta$ or their extensions is feasible).  
Then  the created feasible solution of the size $ U $ is returned.\\

\begin{figure}[H]
	\[\xymatrix @R=3mm {
\textbf{Size}	&	&	&	& *++[F]\txt{$\sigma^\prime$} \ar@{-}[llldd] \ar@{-}[lldd] \ar@{-}[dd] \ar@{-}[rrdd] \ar@{-}[rrrdd]& & &\\
	&	&	&	&	&	&	& \\
\txt{$\beta$}	&	*+[F]\txt{$\sigma^1(\beta)$} \ar@{-}[d] & *+[F]\txt{$\sigma^2(\beta)$} \ar@{-}[d] & \txt{$\cdots$}& *+[F]\txt{$\sigma^{j}(\beta)$} \ar@{-}[d] & \txt{$\cdots$} & *+[F]\txt{$\sigma^{k-1}(\beta)$} \ar@{-}[d] & *+[F]\txt{$\sigma^k(\beta)$} \ar@{-}[d]\\		
\txt{$\beta+1$}	&	*+[F]\txt{$\sigma^{1}_{\beta + 1}$} \ar@{-}[d] & *+[F]\txt{$\sigma^{2}_{\beta + 1}$} \ar@{-}[d] &  & *+[F]\txt{$\sigma^{j}_{\beta + 1}$} \ar@{-}[d] &  & *+[F]\txt{$\sigma^{k-1}_{\beta + 1}$} \ar@{-}[d] & *+[F]\txt{$\sigma^{k}_{\beta + 1}$} \ar@{-}[d]\\
	&	\txt{$\vdots$} \ar@{-}[d] & \txt{$\vdots$} \ar@{-}[d] & & \txt{$\vdots$} \ar@{-}[d] & & \txt{$\vdots$} \ar@{-}[d] & \txt{$\vdots$} \ar@{-}[d]\\		
\txt{$ i $}	&	*+[F]\txt{$\sigma^{1}_{i}$} \ar@{-}[d] & *+[F]\txt{$\sigma^{2}_{i}$} \ar@{-}[d] &  & *+[F]\txt{$\sigma^{j}_{i}$} \ar@{-}[d] &  & *+[F]\txt{$\sigma^{k-1}_{i}$} \ar@{-}[d] & *+[F]\txt{$\sigma^{k}_{i}$} \ar@{-}[d]\\		
	&	\txt{$\vdots$} \ar@{-}[d]& \txt{$\vdots$} \ar@{-}[d] & & \txt{$\vdots$} \ar@{-}[d] & & \txt{$\vdots$} \ar@{-}[d] & \txt{$\vdots$} \ar@{-}[d]\\
\txt{$U-1$}	&	*+[F]\txt{$\sigma^{1}_{U-1}$} & *+[F]\txt{$\sigma^{2}_{U-1}$} &  & *+[F]\txt{$\sigma^{j}_{U-1}$} &  & *+[F]\txt{$\sigma^{k-1}_{U-1}$} & *+[F]\txt{$\sigma^{k}_{U-1}$} 
	}\]
\caption{All possible extensions of the base solutions of size $\beta$, where 
$U$ is a current upper bound}\label{fig1}
\end{figure}
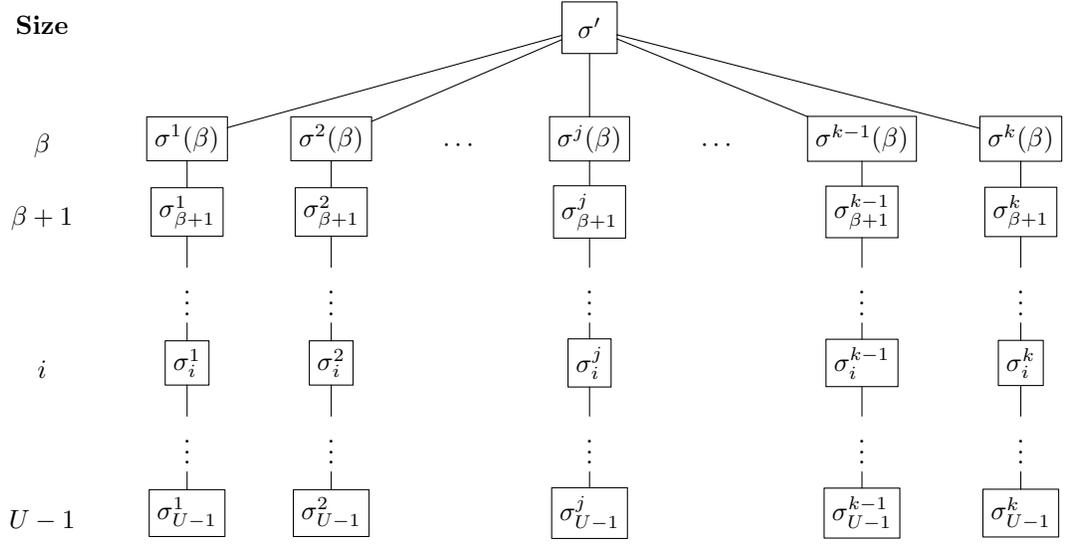

The worst-case behavior of the procedure for a given $\beta$ is
reflected in Figure \ref{fig1}, where the root represents a currently best found 
feasible solution $\sigma^\prime$ and $k=$ 
\begin{scriptsize}
$\left(
\begin{matrix}
	|V(G)\setminus (Supp(G) \cup Leaf(G))|\\
	\beta-s
\end{matrix}
\right)$
\end{scriptsize}.
Note that all the represented solutions must have been infeasible, except 
 $\sigma^\prime$ and possibly $\sigma^k_{U-1}$.


\medskip

An extension $ \sigma_i^h $ of a base solution  $ \sigma^h(\beta) $ of size $i$,
$ \beta < i < U $, is $ \sigma_i^h = \sigma_{i-1}^h \cup x $, where $ x \in V \setminus (Leaf(G) \cup \sigma_{i-1}^h) $ is determined as follows. If solution 
$ \sigma^h(\beta) $ was formed by rule \textit{(1)}, vertex $ x $ is selected randomly; if solution $ \sigma^h(\beta) $ was formed by either of the rules \textit{(2)} and \textit{(3)}, $ x $ is set to be a vertex with the maximum active degree from set $ V \setminus (Leaf(G) \cup \sigma_{i-1}^h) $. A formal 
description of the procedure is as follows:

\medskip

\begin{algorithm}[H]
	\fontsize{10pt}{.3cm}\selectfont
	\caption{$Procedure$\_$Extensions(\sigma^h(\beta))$}\label{alg_DBS_ext}
	\begin{algorithmic}
		\State \textbf{Input:} $\sigma^h(\beta)$, $U$, $Leaf(G)$ .
		\State \textbf{Output:} A feasible solution $\sigma_i^h(\beta)$ or $ NIL $.
		
		\State $i:=|\sigma^h(\beta)|$;
		\State $\sigma_i^h:=\sigma^h(\beta)$;
		\While {$i<U-1$} \hspace{.3cm} \{ Generate new extension of $\sigma^h(\beta)$ \}
			\State $i:=i+1$;
			\If{$h==0$ }
			\State $x:=$ randomly selected vertex from $V \setminus ( Leaf(G) \cup
			 \sigma_{i-1}^h)$;
			\Else
			\State $x:=$ any vertex with the maximum active degree from set $ V \setminus ( Leaf(G) \cup \sigma_{i-1}^h) $;
			\EndIf
			
			\State $ \sigma_i^h = \sigma_{i-1}^h \cup x $;
			\If{$\sigma_i^h $ is a feasible solution}
				\State return $\sigma_i^h $ and stop;  
			\EndIf
		\EndWhile
		\State return $NIL $;

	\end{algorithmic}
\end{algorithm} 

\medskip

Now we give a description of the overall heuristic Algorithm \ref{alg_DBS} that combines depth-first search with a breadth-first search (while the base solutions are generated by the breadth-first rule, the extensions of every base solution are generated in depth-first fashion, see Figure \ref{fig1}): 

\begin{algorithm}[H]
	\fontsize{8pt}{.25cm}\selectfont
	\caption{Algorithm\_DBS}\label{alg_DBS}
	\begin{algorithmic}
		\State \textbf{Input:} A graph $G$ of order $n$.
		\State \textbf{Output:} A feasible solution  $ S^* $. 
						
		\State Generate feasible solution $\sigma$;
		\State $U := |\sigma|$;
		\State $s := |Supp(G)|$;
		\State $l := |Leaf(G)|$;
		\State $\alpha =$ value random between $ 0.2 $ and $ 0.7 $;
		\State $  \beta = \lfloor\alpha (U-s)\rfloor + s $;  
		
		\{ Iterative step \} 
		\State $h:=0$;

		\State $	h_{max}:=\left(
			\begin{matrix}
				n-s-l\\
				\beta-s
			\end{matrix}
			\right)$;	
		
		\While{$h<h_{max}$}
		 
		\{ Generate new base solution $\sigma^h(\beta)$ \}
		\If{$h==0$ }
		\State $\sigma^h(\beta) := Supp(G)\cup\{$ the first $  \beta-s $ vertices from solution $ \sigma\setminus Supp(G)\}$  ;
		\Else
		\If{$h<\left(
			\begin{matrix}
				|\sigma|-s\\
				\beta-s
			\end{matrix}
			\right)$ }  
		\State $\sigma^h(\beta) := Supp(G)\cup\{$ randomly selected $  \beta-s $ vertices from solution $ \sigma\setminus Supp(G)\}$;
		\Else
		\State $\sigma^h(\beta) := Supp(G)\cup\{$ randomly selected $  \beta-s $ vertices from $ V(G)\setminus ( Supp(G)\cup Leaf(G))\}$;
		\EndIf 
		\EndIf
		
		\If{$\sigma^h(\beta) $ is a feasible solution}
			\State $\sigma:=\sigma^h(\beta)$;
			\State $U := |\sigma|$;
			\State $  \beta = \lfloor\alpha (U-s)\rfloor + s $;  
			\State $h:=0$;
			\State $	h_{max}:=\left(
			\begin{matrix}
				|n-s-l|\\
				\beta-s
			\end{matrix}
			\right)$;
		\Else
		
		\{ Generate extensions of $\sigma^h(\beta)$ \}
		\If{$Procedure\_Extensions(\sigma^h(\beta))$ returns $NIL$}
		\State $h:=h+1$;
		\Else  \hspace{.3cm} \{ A feasible solution was found ($\sigma_i^h$)\}
		\State $\sigma:=\sigma_i^h(\beta)$;
		\State $U := |\sigma|$;
		\State $  \beta = \lfloor\alpha (U-s)\rfloor + s $;
		\State $h:=0$;
		\State $	h_{max}:=\left(
					\begin{matrix}
					n-s-l\\
					\beta-s
					\end{matrix}
					\right)$;
		\EndIf
		\EndIf

		\EndWhile
		\State $S^*:=\sigma$;
	\end{algorithmic}
\end{algorithm} 
\medskip

\newpage
The next two observations immediately follow from the construction 
of Algorithm \ref{alg_DBS}: 
\begin{obs}\label{obs5}
	If no extension of any base solution of current size $ \beta $ is feasible and  $ \beta > L $, then $\beta < \gamma(G) \leq U$. 	 
\end{obs}
\begin{obs}\label{obs2}
Algorithm\_DBS inherits the approximation ratio of any algorithm used to find the initial feasible solution $\sigma$. 
\end{obs}
In the next two theorems, we give approximation ratios of Algorithm \ref{alg_DBS}. 
\begin{thm}\label{prop1}
For a  connected graph $G(V,E)$ of maximum degree $\Delta(G)$, Algorithm\_DBS  
gives the following approximation ratio
	\begin{center}
	$\rho \leq \left\{\begin{array}{ll}
		\frac{\Delta + 1}{2}, & \mbox{if 1 $\leq \Delta \leq$ 4} \\[.3cm]
		ln(\Delta+1)+1, & \mbox{otherwise.}
	\end{array}
	\right.$
	\end{center}	 	 
\end{thm}
\begin{proof}
Let $\sigma$ be the feasible solution delivered by the algorithm from
\cite{Mira2021}. If $|\sigma| \leq 2$, then $\sigma$ is a minimum 
dominating set (see Remark 1 from \cite{Mira2021}). Thus assume 
$\gamma(G) > 2$. We define \emph{private neighborhood} of vertex 
$v \in \sigma\subseteq V$, as $\{u \in V: N(u)\cap \sigma = \{v\}\}$; 
a vertex in the private neighborhood of 
vertex $v$ is said to be its \emph{private neighbor} with respect to set $\sigma$.
Since $G$ is a connected graph, all $x\in \sigma$ has at least one private neighbor 
(see \cite{Mira2021}). Hence,  $|\sigma| \leq n/2$. By Theorem \ref{teorema1}, 
$\gamma(G) \geq \frac{n}{\Delta+1}$.  
Then $\frac{n}{\Delta+1} \leq |\sigma| \leq \frac{n}{2}$. Since $|\sigma|\geq \gamma(G)$, we obtain the following expression for an approximation ratio 
$\displaystyle\rho = \frac{|\sigma|}{\gamma(G)} \leq \frac{\frac{n}{2}}{\frac{n}{\Delta+1}} = \frac{\Delta+1}{2}$. Another valid ratio is 
$\rho \leq  ln(\Delta+1)+1$ since it is one of the algorithm in 
\cite{Mira2021}.
\end{proof}
\begin{thm}\label{prop2}
	Let $G(V,E)$ be a connected graph of diameter $d$ and radius $r$. Then
	Algorithm\_DBS  gives the following approximation ratio
	$$\rho=\min\left\lbrace \frac{3n}{2(d+1)},\frac{3n}{4r}\right\rbrace . $$
\end{thm}
\begin{proof}
By Theorem \ref{teorema2}, $\gamma(G) \geq \frac{d+1}{3}$ and $\gamma(G) \geq \frac{2r}{3}$. At the same time,   $|\sigma| \leq n/2$ (see the above proof). 
Then $\frac{d+1}{3} \leq |\sigma| \leq \frac{n}{2}$ and $\frac{2r}{3} \leq |\sigma| \leq \frac{n}{2}$. Since $|\sigma|\geq \gamma(G)$, we obtain the following
approximation ratios  
$\displaystyle\rho_1 = \frac{|\sigma|}{\gamma(G)} \leq \frac{\frac{n}{2}}{\frac{d+1}{3}} = \frac{3n}{2(d+1)}$ and 
$\displaystyle\rho_2 = \frac{|\sigma|}{\gamma(G)} \leq \frac{\frac{n}{2}}{\frac{2r}{3}} = \frac{3n}{4r}$. Hence, $\rho=\min\left\lbrace  \rho_1,\rho_2   \right\rbrace$
is another valid approximation ratio.  
\end{proof}

\section{Experimental results}

In this section, we describe our computational experiments. We implemented the three algorithms in C++ using Visual Studio for Windows 10 (64 bits) on a personal computer with Intel Core i7-9750H (2.6 GHz) and 12 GB of RAM DDR4.  The order and the size of an instance were generated randomly using function \textit{random()}. To complete the set E(G), each new edge was added in between two yet non-adjacent vertices randomly until the corresponding size was reached.\\

Throughout this section, the exact algorithms from \cite{van2011exact}, \cite{iwata2012faster}, the ILP formulation from Section 3, and our exact 
algorithm will be refereed to as $ MSC^1 $, $ MSC^2 $, ILP and BDS, 
respectively, and our heuristic will be refereed to as DBS.
We analyzed  more than  1100 randomly generated instances which are
publicly available at  \cite{bdparra1} and \cite{bdparra2}. A complete 
summary of the results for these instances can be found at \cite{bdparra2}.
We choose  72 from these instances randomly and report the results for them. \\

\begin{figure}[H]
	\centering
	\begin{subfigure}[b]{0.6\linewidth}
		\includegraphics[width=\linewidth]{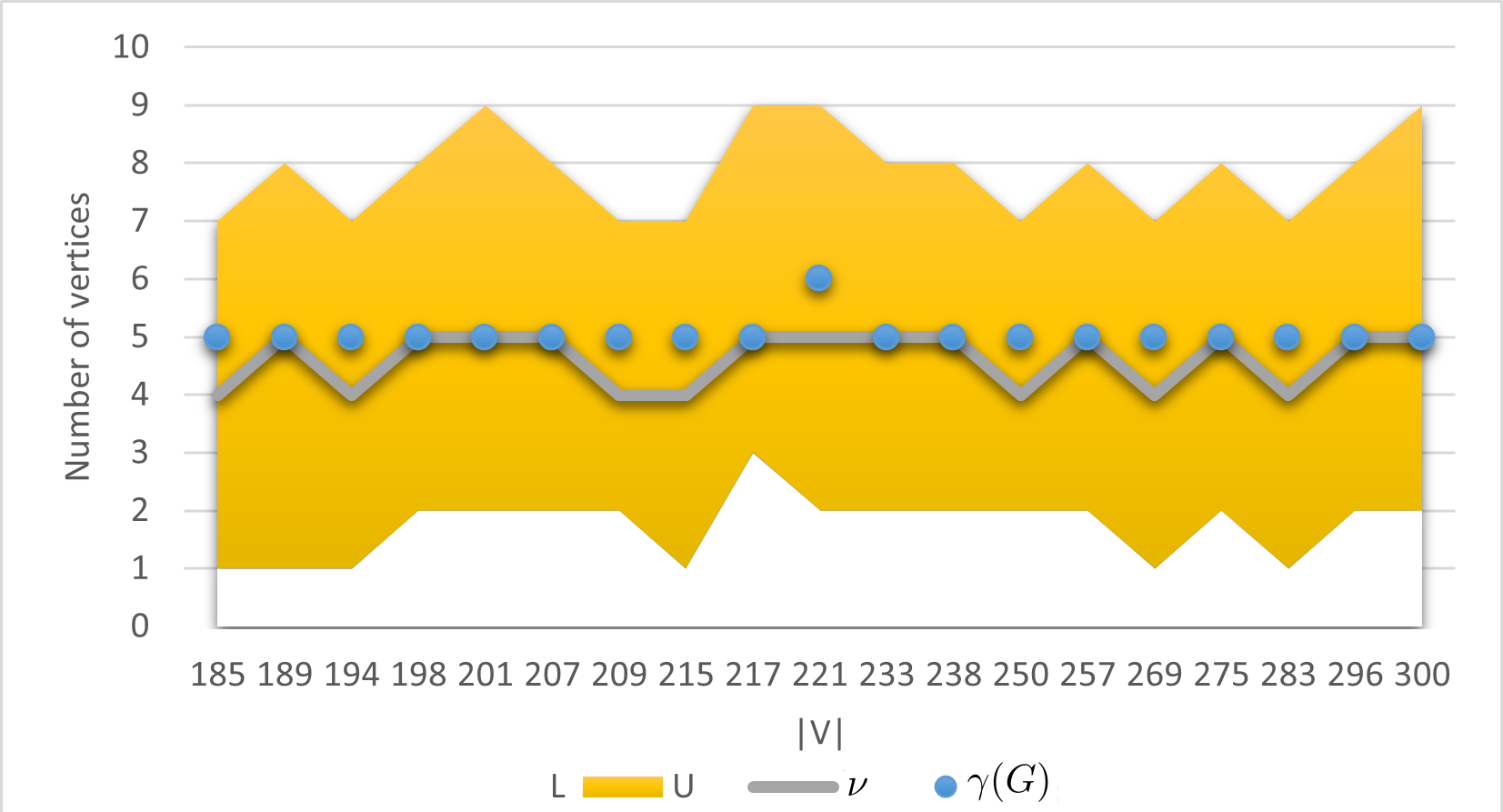}
		\caption{Iteration 1}
		\label{figura7}
	\end{subfigure}
	\begin{subfigure}[b]{0.6\linewidth}
		\includegraphics[width=\linewidth]{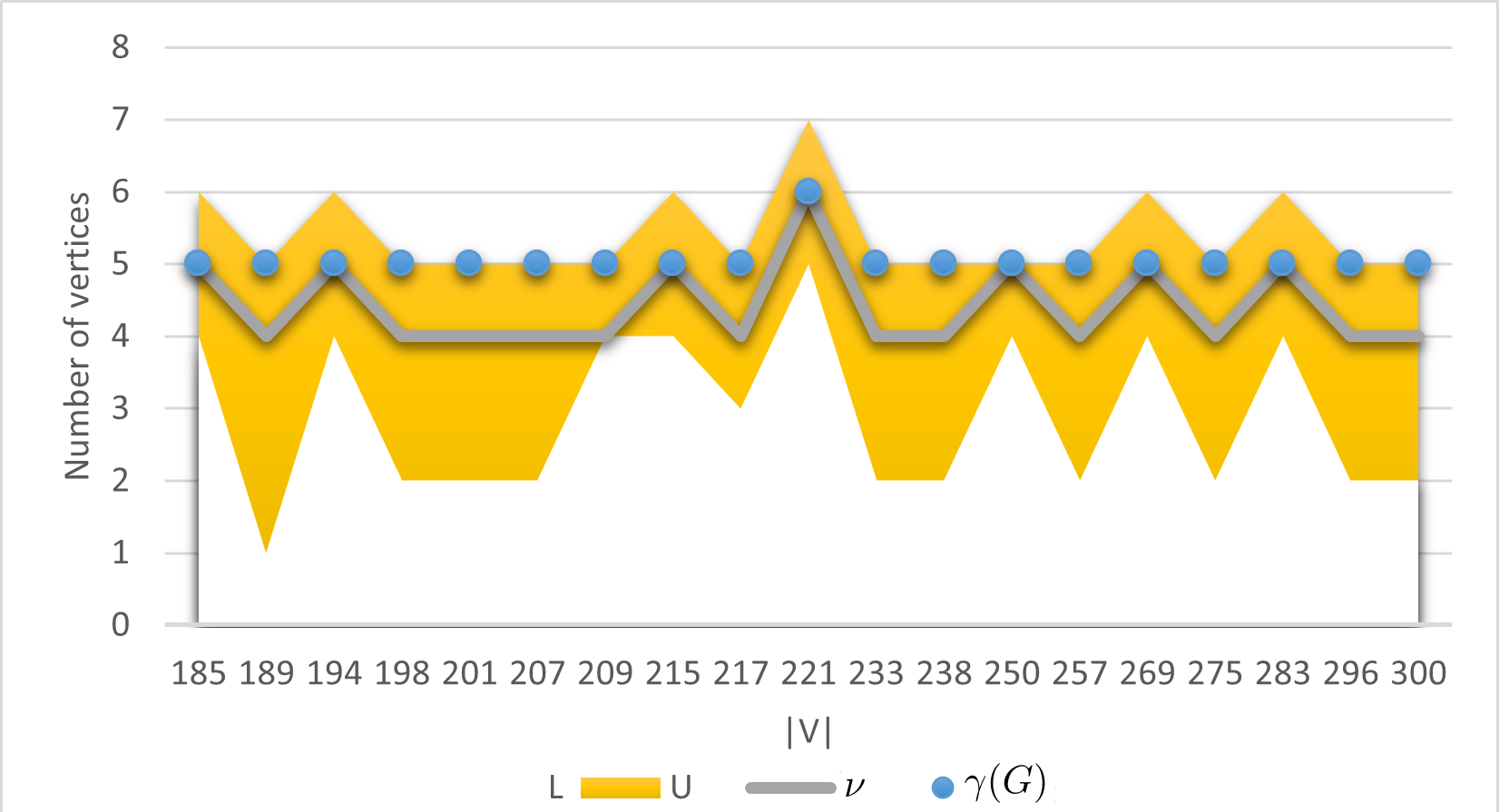}
		\caption{Iteration 2}
		\label{figura8}
	\end{subfigure}
	\begin{subfigure}[b]{0.6\linewidth}
		\includegraphics[width=\linewidth]{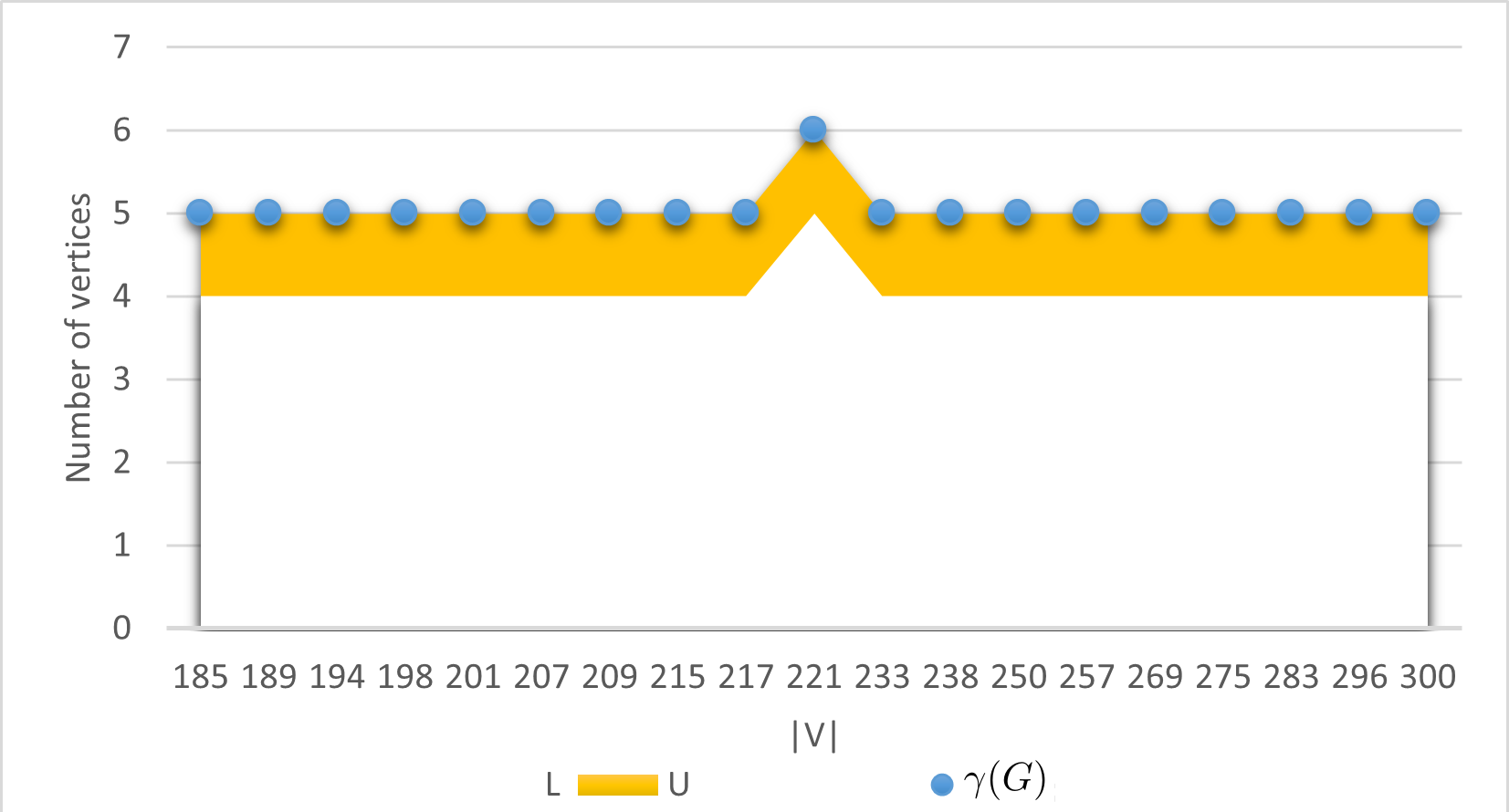}
		\caption{Iteration 3}
		\label{figura9}
	\end{subfigure}
	\caption{Lower and upper bounds of the search interval for BDS-Algorithm at three consecutive iterations, and $ \gamma(G)$}
	\label{figura7_9}
\end{figure}

The input for BDS-Algorithm is a feasible solution that can be obtained by any
heuristic. Since the results of the two-stage algorithm from \cite{Mira2021}
gives the best approximation that we know, we choose this particular
algorithm to create the inputs of BDS-Algorithm  (the description of the solutions created by the former algorithm can be found in \cite{bdparra1}). In Figure \ref{figura7_9} we represent how lower and upper bounds are 
updated during the first three iterations of BDS-Algorithm and how the
number of nodes in the generated feasible solutions is reduced step-by-step
for 19 middle-sized middle-dense instances.

Table \ref{tabla5} presents the results for small-sized
quasi-trees with the density of approximately 0.2, where ILP formulation 
resulted in the best performance (due to small amount of variables in linear 
programs). The two MSC-Algorithms 
turned out to be faster than BDS-Algorithm (since the vertex sets used in the
former two algorithms are very small and the accomplished reductions have 
low computational costs).

\begin{table}[H] 
	\centering
	\resizebox{12.5cm}{!}{%
		\begin{tabular}{cccccccccccccc}
			\hline
			\multirow{2}{*}{No.} &
			\multirow{2}{*}{$|V(G)|$} &
			\multirow{2}{*}{$|E(G)|$} &
			\multicolumn{4}{c}{Time}&
			\multicolumn{4}{c}{Lower Bounds} &
			\multirow{2}{*}{$\gamma(G)$} &
			\multicolumn{2}{c}{Upper Bounds} \\ \cline{4-11} \cline{13-14} 
			&
			&
			&
			ILP&
			BDS &
			$ MSC^1 $ &
			$ MSC^2 $&
			$\frac{n}{\Delta(G)+1}$ &
			$\frac{d+1}{3}$ &
			$\frac{2r}{3}$ &
			$|Supp(G)|$ &
			&
			$|S|$ &
			$n-\Delta(G)$ \\ 
			\hline
			
			1  & 50  & 286  & 0.16 & 1.89 & 0.61  & 0.23 & 2                       & 1               & 1              & 2        & 6 & 6      & 33                       \\ 
			2  & 60  & 357  & 0.11 & 3.13 & 0.71  & 0.29 & 2                       & 1               & 1              & 1        & 6 & 7      & 40                        \\ 
			3  & 70  & 524  & 0.12 & 7.04 & 1.46  & 0.37 & 2                       & 1               & 1              & 1        & 6 & 6      & 45                        \\ 
			4  & 80  & 678  & 0.13 & 12.59 & 2.81  & 0.79 & 2                       & 1               & 1              & 1        & 6 & 7      & 53                        \\ 
			5  & 90  & 842  & 0.17 & 390.90 & 5.18  & 2.31 & 3                       & 1               & 1              & 1        & 7 & 8      & 63                        \\ 
			6  & 100 & 1031 & 0.21 & 803.74 & 8.31  & 3.01 & 2                       & 1               & 1              & 2        & 7 & 7      & 67                       \\ 
			7  & 101 & 1051  & 2.17 & 804.20 & 8.69  & 3.17 & 3                       & 1               & 1              & 1        & 7 & 7      & 70                        \\ 
			8  & 106 & 1154  & 1.78 & 1143.92 & 10.82  & 2.93 & 2                       & 1               & 1              & 1        & 7 & 8      & 71                        \\ 
			9  & 113 & 1306  & 1.29 & 1594.92 & 16.25  & 4.27 & 3                       & 1               & 1              & 1        & 7 & 7      & 77                      \\ 
			10 & 117 & 1398  & 2.56 & 3364.4  & 19.95  & 5.33 & 3                       & 1               & 1              & 2        & 8 & 9      & 84                        \\ 
			11 & 121 & 1493  & 3.22 & 2240.49 & 23.11  & 4.86 & 3                       & 1               & 1              & 1        & 7 & 7      & 87                        \\
			\hline
	\end{tabular}}
	\caption{The results for the graphs with density $\approx 0.2$.\label{tabla5}}
\end{table}

For middle-dense graphs where {\it all four} exact algorithms succeeded to halt with an optimal solution,  BDS  and ILP algorithms provided better execution times than MSC-Algorithms;  ILP formulation resulted in worse times than the $MSC^2$-Algorithm for instances with more than 295 vertices, and  
BDS-Algorithm was faster than ILP formulation for instances from  255 
vertices, see Table \ref{tabla1} and Figure \ref{figura2} below.

\begin{table}[H] 
	\centering
	\resizebox{12.5cm}{!}{%
	\begin{tabular}[H]{cccccccccccccc}
		\hline
		\multirow{2}{*}{No.} &
		\multirow{2}{*}{$|V(G)|$} &
		\multirow{2}{*}{$|E(G)|$} &
		\multicolumn{4}{c}{Time} &
		\multicolumn{4}{c}{Lower Bounds} &
		\multirow{2}{*}{$\gamma(G)$} &
		\multicolumn{2}{c}{Upper Bounds} \\\cline{4-7} \cline{8-11} \cline{13-14} 
		&
		&
		&
		 ILP  &
		 BDS  &
		$ MSC^1 $ &
		$ MSC^2 $&
		$\frac{n}{\Delta(G)+1}$ &
		$\frac{d+1}{3}$ &
		$\frac{2r}{3}$ &
		$|Supp(G)|$ &
		&
		$|S|$ &
		$n-\Delta(G)$ \\
		\hline 
		
		%
		
		1 & 185 & 6849 & 2.75 & 24.66    & 666.83 & 176.18 & 1 & 1 & 1 & 1 & 5 & 6 & 93    \\ 
		2 & 189 & 7147 & 3.35 & 23.85   & 770.36  & 206.02 & 1 & 1 & 1 & 1 & 5 & 7 & 91    \\ 
		3 & 194 & 7529 & 3.29 & 28.29    & 915.36 & 335.88 & 1 & 1 & 1 & 1 & 5 & 6 & 96    \\ 
		4 & 198 & 7842 & 3.61 & 30.53   & 1041.98 & 395.03 & 2 & 1 & 1 & 1 & 5 & 6 & 101   \\ 
		5 & 201 & 8081 & 2.78 & 31.03   & 1136.67 & 313.15 & 2 & 1 & 1 & 2 & 5 & 7 & 103  \\ 
		6 & 207 & 8569 & 3.28  & 35.94   & 1361.77 & 659.35 & 2 & 1 & 1 & 1 & 5 & 6 & 105  \\ 
		7 & 209 & 8735 & 6.40  & 37.35   & 1454.44 & 715.74 & 1 & 2 & 1 & 1 & 5 & 5 & 104  \\ 
		8 & 215 & 9243 & 8.25  & 42.91   & 1810.09 & 499.96 & 1 & 1 & 1 & 1 & 5 & 6 & 103  \\ 
		9 & 217 & 9415 & 9.16  & 42.88    & 1892.35 & 890.39 & 1 & 1 & 1 & 3 & 5 & 6 & 107  \\ 
		10 & 221 & 9765& 13.73   & 49.91     & 2011.15 & 866.37 & 2 & 1 & 1 & 1 & 6 & 7 & 115  \\ 
		11 & 233 & 10852& 12.74  & 57.83   & 2793.59 & 1028.85 & 2 & 1 & 1 & 1 & 5 & 6 & 118  \\ 
		12 & 238 & 11322& 53.59  & 65.05    & 3093.64 & 1485.16 & 2 & 1 & 1 & 1 & 5 & 6 & 123  \\ 
		13 & 250 & 12491& 66.09  & 83.19   & 4065.44 & 1988.04 & 1 & 2 & 1 & 1 & 5 & 5 & 125  \\ 
		14 & 257 & 13199& 118.24  & 93.98    & 4919.85 & 1478.89 & 2 & 1 & 1 & 1 & 5 & 6 & 131  \\ 
		15 & 269 & 14459& 104.05  & 112.609   & 6102.11 & 2677.25 & 1 & 1 & 1 & 1 & 5 & 6 & 134  \\ 
		16 & 275 & 15111& 143.66  & 114.446   & 6887.98 & 3082.72 & 2 & 1 & 1 & 1 & 5 & 6 & 144  \\ 
		17 & 283 & 16002& 163.58  & 418.72    & 8040.4  & 3649.57 & 1 & 1 & 1 & 1 & 5 & 6 & 141  \\ 
		18 & 296 & 17505& 6071.5  & 155.861   & 10269.4 & 4467.34 & 2 & 1 & 1 & 1 & 5 & 6 & 152  \\ 
		19 & 300 & 17981& 6318.52  & 165.301   & 11035.8 & 4811.58 & 2 & 1 & 1 & 1 & 5 & 7 & 151  \\ 
		\hline
		
	\end{tabular}}
	\caption{The results for the graphs with density $\approx 0.5$.\label{tabla1}}
\end{table}

\begin{figure}[H]
	\centering
	\includegraphics[width=1.0\linewidth]{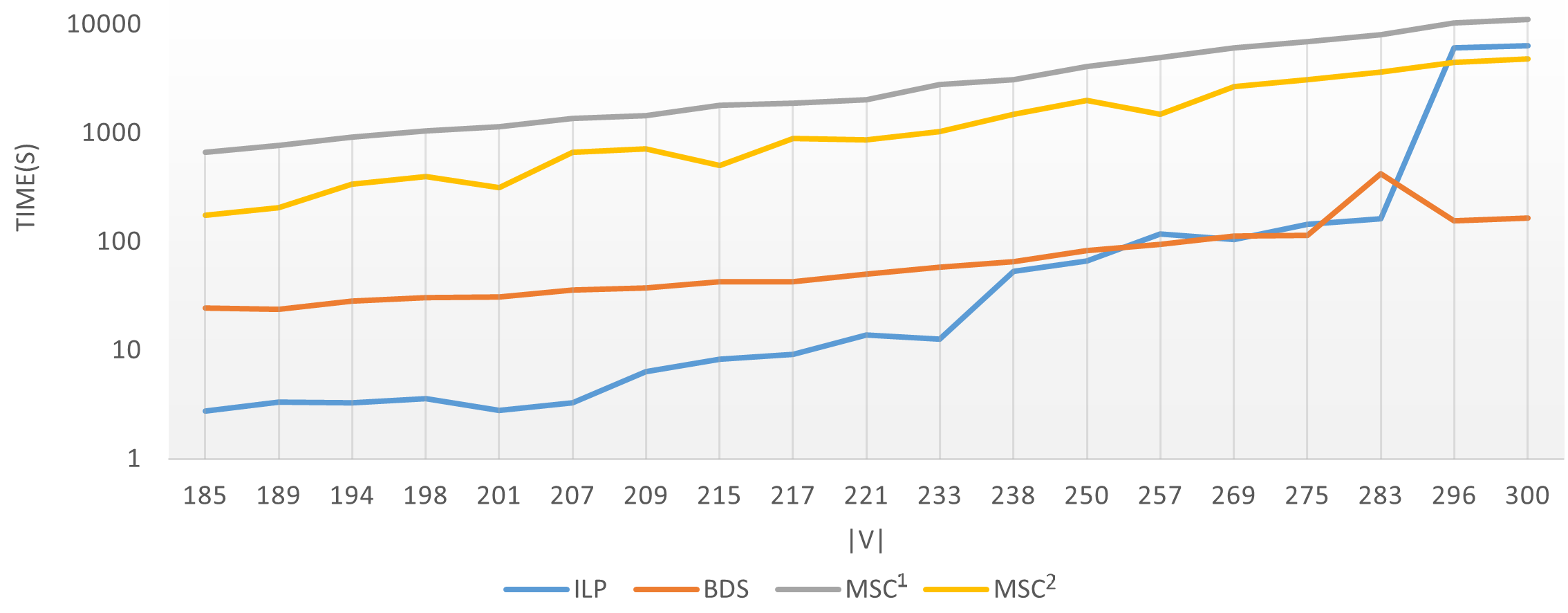}
	\caption[time]{Execution times for the four exact algorithms
	(logarithmic scale was applied).}
	\label{figura2}
\end{figure}

\begin{figure}[H]
	\centering
	\includegraphics[width=1.0\linewidth]{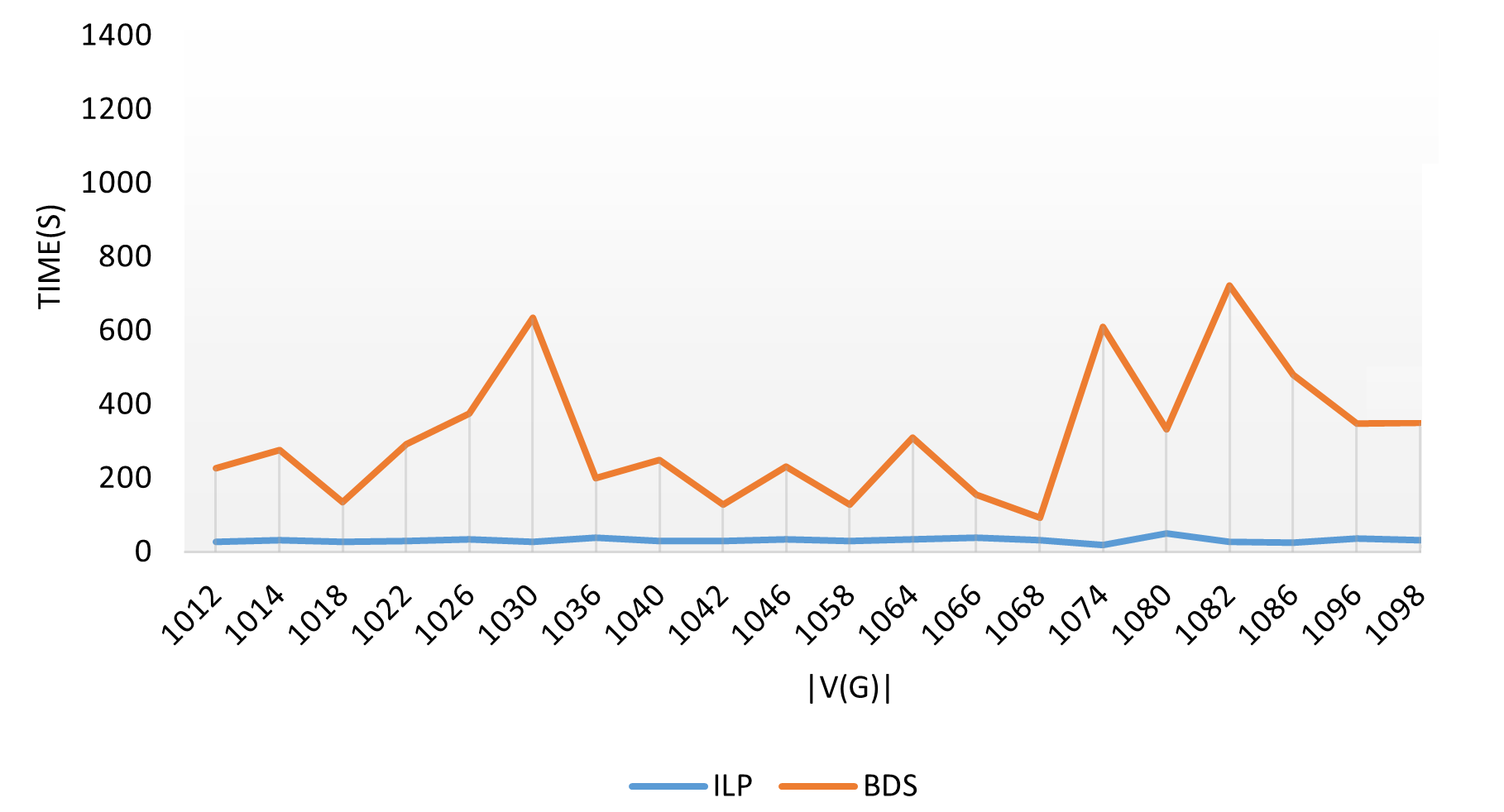}
	\caption{BDS vs. ILP for dense graphs}
	\label{figura3}
\end{figure}

Table \ref{tabla3} presents results for larger instances with the density 
$\approx 0.8$, where MSC-Algorithms failed to halt within  8 hours of 
time limit; algorithms BDS and ILP completed within 6 minutes for the largest instance with 1098 vertices. These instances are dense, and ILP formulation results in smaller execution times (see Figure \ref{figura3}).  These dense graphs are quasi complete graphs, so $\gamma(G)$ is very small ranging
from 4 to 5.

\begin{table}[H] 
	\centering
	\resizebox{12.5cm}{!}{%
	\begin{tabular}[c]{cccccccccccc}
		\hline
		\multirow{2}{*}{No.} & \multirow{2}{*}{$|V|$} & \multirow{2}{*}{$|E|$}& \multicolumn{2}{c}{Time} & \multicolumn{4}{c}{Lower Bounds} & \multirow{2}{*}{$\gamma(G)$} & \multicolumn{2}{c}{Upper Bounds} \\ \cline{4-9} \cline{11-12} 
		&  &  & ILP & BDS & $\frac{n}{\Delta(G)+1}$ & $\frac{d+1}{3}$ & $\frac{2r}{3}$ & $|Supp(G)|$ &  & $|S|$ & $n-\Delta(G)$ \\ \hline
		%
		%
		%
		1	& 1012 & 452294 & 26.27 & 227.34 & 1 & 2 & 2 & 2 & 5 & 6 & 76 \\
		2	& 1014 & 454116 & 31.72 & 275.10 & 1 & 2 & 2 & 2 & 5 & 6 & 79 \\
		3	& 1018 & 406867 & 27.23 & 135.86 & 1 & 2 & 2 & 2 & 5 & 7 & 175 \\
		4	& 1022 & 461340 & 29.27 & 291.41 & 1 & 2 & 2 & 2 & 5 & 6 & 80 \\
		5	& 1026 & 413355 & 34.74 & 375.12 & 1 & 2 & 2 & 2 & 5 & 7 & 176 \\
		6	& 1030 & 416585 & 27.07 & 633.91 & 1 & 2 & 2 & 2 & 5 & 7 & 164 \\
		7	& 1036 & 474219 & 38.03 & 198.76 & 1 & 2 & 2 & 2 & 5 & 6 & 86 \\
		8	& 1040 & 477909 & 29.73 & 248.36 & 1 & 2 & 2 & 2 & 5 & 6 & 83 \\
		9	& 1042 & 426451 & 28.48 & 127.95 & 1 & 2 & 2 & 2 & 5 & 7 & 174 \\
		10	& 1046 & 483492 & 32.97 & 231.50 & 1 & 2 & 2 & 2 & 5 & 6 & 78 \\
		11	& 1058 & 439752 & 29.10 & 128.97 & 1 & 2 & 2 & 2 & 5 & 7 & 162 \\
		12	& 1064 & 500420 & 33.63 & 310.40 & 1 & 2 & 2 & 2 & 5 & 6 & 83 \\
		13	& 1066 & 446505 & 37.56 & 155.74 & 1 & 2 & 2 & 2 & 5 & 6 & 173 \\
		14	& 1068 & 504205 & 31.27 & 92.06 & 1 & 2 & 2 & 2 & 4 & 6 & 81 \\
		15	& 1074 & 453307 & 19.43 & 610.30 & 1 & 2 & 2 & 2 & 5 & 7 & 174 \\
		16	& 1080 & 515722 & 49.56 & 333.09 & 1 & 2 & 2 & 2 & 5 & 6 & 79 \\
		17	& 1082 & 460117 & 26.83 & 723.12 & 1 & 2 & 2 & 2 & 5 & 7 & 183 \\
		18	& 1086 & 463560 & 24.05 & 480.90 & 1 & 2 & 2 & 2 & 5 & 7 & 182 \\
		19	& 1096 & 531243 & 35.54 & 347.71 & 1 & 2 & 2 & 2 & 5 & 6 & 87 \\
		20	& 1098 & 533220 & 31.33 & 349.20 & 1 & 2 & 2 & 2 & 5 & 6 & 81 \\ 
		\hline
	\end{tabular}}
		\caption{Results for the graphs with density $\approx 0.8$}
		\label{tabla3}
	\end{table}

In Table 4 below, $S$ is the feasible solution from \cite{Mira2021}. $S_h^*$ is the feasible solution of iteration $h$, column $\sigma^j(\beta)$ is the number of base solutions generated at iteration $h$ , and column ``Time''  is for running times in seconds, all these parameters are of DBS-Algorithm. For all
instances in Table 4, DBS-Algorithm proceeded in at most 3 external iterations
so that no feasible solution was found either at the second or at the third
iteration. In Table 4, the blank spaces in the column ``Iteration 2'' indicate 
that no feasible solution at iteration 2 was found (the feasible solution
of iteration 1 could have been already optimal). \\

\begin{figure}[H]
	\centering
	\includegraphics[width=0.85\linewidth]{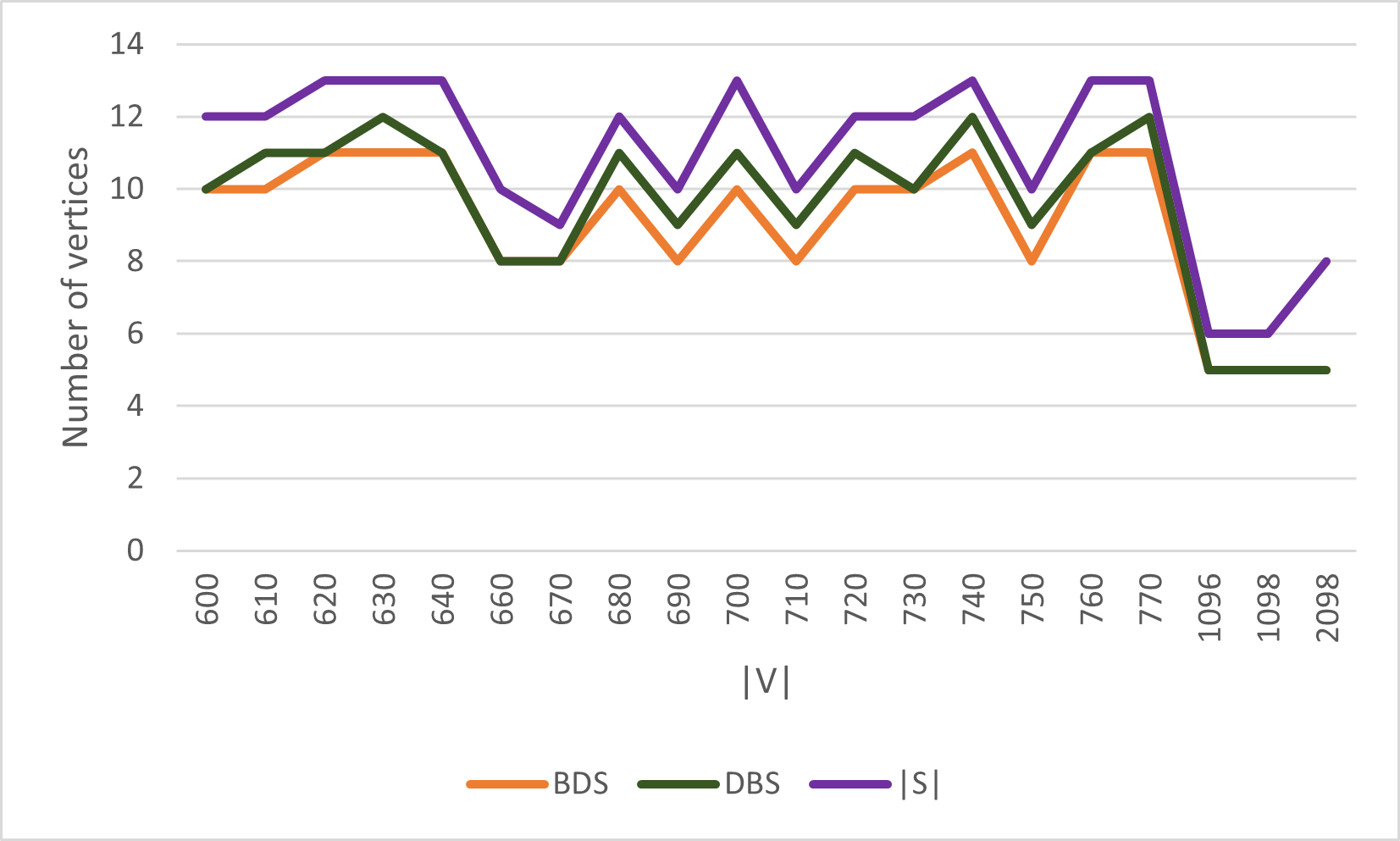}
	\caption[time]{Approximation results}
	\label{figura51}
\end{figure}

Table \ref{tabla4} and Figure \ref{figura51} illustrate
comparative quality of the solutions generated by algorithms \cite{Mira2021} 
and DBS for the large-sized instances with up to 2098 vertices. Remarkably,
in 98.62\% of the instances, the solutions from \cite{Mira2021} were improved 
by DBS-Algorithm. The latter algorithm found an optimal solution for 61.54\% of 
the analyzed instances, whereas for the remaining instances, the average 
approximation error was $ 1.18 $.

\begin{figure}[H]
	\centering
	\includegraphics[width=0.85\linewidth]{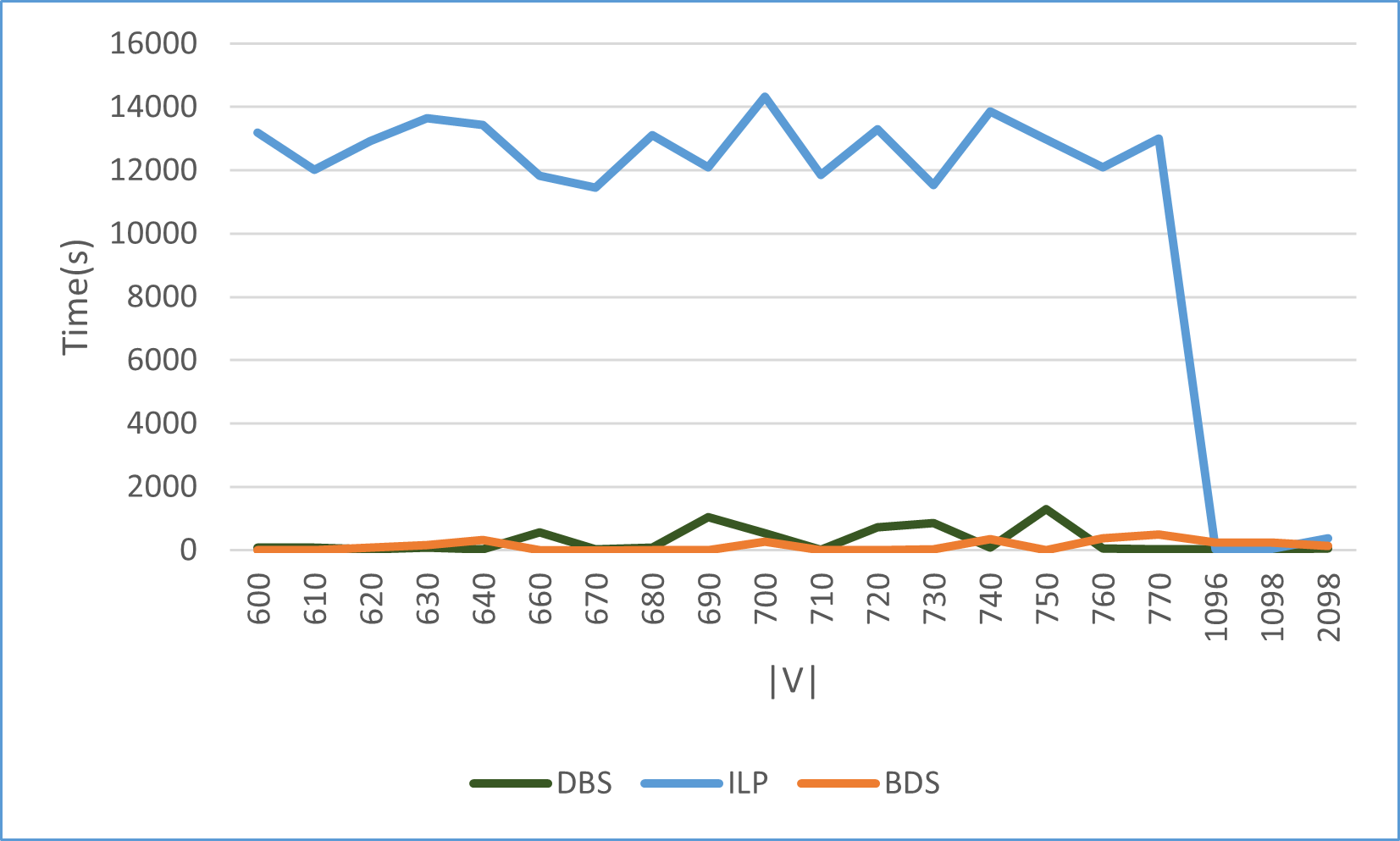}
	\caption[time]{Time lapse for finding optimal and sub-optimal solutions 
   by BDS, ILP and DBS }
	\label{figura61}
\end{figure}

As it can be observed  from Table \ref{tabla4} and Figure 
\ref{figura61}, BDS and DBS algorithms find optimal solutions faster than ILP 
does, whereas the solutions generated by DBS-algorithm give optimal or very close 
to the optimum value. We can also observe a sharp improvement of
ILP for the last three dense instances (with the density more than 0.8). In average, 
DBS-Algorithm was 243 times faster than the formulation ILP, providing 
optimal or very close to the optimal solutions. BDS-Algorithm was able to
find optimal solutions, in average, 874 times faster than  ILP. We may
observe that there are instances where DBS-Algorithm is slower than 
BDS-Algorithm. This happened due to the large number of base solutions and 
their extensions created by DBS-Algorithm, whereas for these instances, 
BDS-Algorithm enumerated considerably less solutions before finding 
an optimal one: in BDS-Algorithm, whenever  $\nu \geq \gamma(G)$, a feasible
solution of cardinality $\nu$ was normally found very fast due to the
order in which the solutions of a given size are enumerated in this 
algorithm (see priority lists in Section 4.1).\\

\begin{table}[H] 
	\centering
	\resizebox{12.5cm}{!}{%
	\begin{tabular}{ccccccccccccccc}
		\hline
		\multicolumn{1}{l}{\multirow{2}{*}{No.}} &
		\multicolumn{1}{l}{\multirow{2}{*}{$|V(G)|$}} &
		\multicolumn{1}{l}{\multirow{2}{*}{$|E(G)|$}} &
		\multicolumn{1}{l}{\multirow{2}{*}{$|S|$}} &
		\multicolumn{4}{c}{Iteration 1} &
		\multicolumn{4}{c}{Iteration 2} & 
		\multicolumn{1}{l}{\multirow{2}{*}{$\gamma(G)$}}&\multicolumn{2}{c}{Time} \\ \cline{5-12} \cline{14-15}
		\multicolumn{1}{l}{} &
		\multicolumn{1}{l}{} &
		\multicolumn{1}{l}{} &
		\multicolumn{1}{l}{} &
		\multicolumn{1}{l}{$\beta$} &
		\multicolumn{1}{l}{$\sigma^j(\beta)$} &
		\multicolumn{1}{l}{Time} &
		\multicolumn{1}{l}{$|S_1^*|$} &
		$\beta$ &
		\multicolumn{1}{l}{$\sigma^j(\beta)$} &
		\multicolumn{1}{l}{Time} &
		\multicolumn{1}{l}{$|S_2^*|$} & &
		\multicolumn{1}{l}{ILP} &
		\multicolumn{1}{l}{BDS} 
		\\   
		\hline
		
		
		1  & 600 & 84557  & 12 & 4 & 51    & 87.15  & 10 &   &                      &         &   &10 & 13192.35 & 11.05 \\
		2  & 610 & 87490  & 12 & 4 & 102  & 86.61 & 11 &   &                      &         &   &10 & 12007.21 & 13.09 \\
		3  & 620 & 90472  & 13 & 3 & 21    & 20.95  & 12 & 3 & 6                    & 25.74  & 11&11 & 12939.63 & 84.58 \\
		4  & 630 & 93505  & 13 & 4 & 107   & 90.53  & 12 & 3 & 35266                & 29750   & 11&11 & 13662.49 & 176.09 \\
		5  & 640 & 96587  & 13 & 4 & 22    & 23.20  & 11 &   &                      &         &   &11 & 13438.97 & 324.92 \\
		6  & 650 & 102571 & 9  & 2 &       &        & 9  &   &                      &       &       &8 & 11502.59 & 7.69 \\
		7  & 660 & 105798 & 10 & 3 & 280  & 553.46 & 8  &   &                      &         &   &8 & 11837.33 & 9.53 \\
		8  & 670 & 109076 & 9  & 2 & 18    & 24.96  & 8  &   &                      &         &   &8 & 11461.37  & 2.82 \\
		9  & 680 & 109417 & 12 & 4 & 65    & 86.75  & 11 &   &                      &         &   &10 & 13116.93  & 14.09 \\
		10 & 690 & 117488 & 10 & 3 & 812   & 1055.51 & 9  &   &                      &         &   &8 & 12094.02  & 13.18 \\
		11 & 700 & 116132 & 13 & 4 & 384   & 652.76  & 11 &   &                      &         &   &10 &14332.51  & 283.76 \\
		12 & 710 & 120941 & 10 & 3 & 12    & 21.20  & 9  &   &  &         &                       &8 & 11853.09 & 3.37 \\
		13 & 720 & 127996 & 12  & 2 & 398 & 715.17 & 11  &   &  &         &                       &10 & 13305.55 & 14.56 \\
		14 & 730 & 131249 & 12  & 3 & 584 & 853.31 & 10  &   &  &         &                       &10 & 11539.34 & 27.61 \\
		15 & 740 & 130162 & 13 & 4 & 52    & 72.59  & 12 & 3 & 31056                & 40681.6 & 11&11 & 13855.82 & 360.54 \\
		16 & 750 & 137096 & 10 & 3 & 948  & 1305.18 & 9  &   &  &         &                       &8 & 12971.74 & 4.39\\
		17 & 760 & 123941 & 13 & 4 & 19     & 58.63  & 11  &   &                      &         &   &11 & 12109.04 &373.16\\
		18 & 770 & 141210 & 13 & 4 & 13    & 24.70  & 12 & 3 & 9561                 & 16496.5 & 11&11 & 13009.37 & 501.14\\
		19 & 1096 & 531243 & 6 & 2 & 3    & 25.66  & 5 &   &                   &   &              &5 & 35.54 & 347.71\\
		20 & 1098 & 533220 & 6 & 2 & 4    & 28.11  & 5 &   &                   &   &              &5 & 31.33 & 349.2\\
		21 &  2098 & 1962869 & 8 & 3 &  14   &  58.22 & 5 &   &                   &   &           &5 & 387.78 & 104.81 \\
		
		\hline
	\end{tabular}}
\caption{Performance of DBS vs BDS and ILP \label{tabla4}}
\end{table}

\section{Conclusions}

Different exact and heuristic algorithms for a variety of graph 
domination problems can be constructed based on the ideas presented here (such
problems include global dominant, $k$-dominant, total dominant, global total and 
global total $k$-dominant settings). It can be beneficial to use
different search strategies for different types of graphs. For instance, 
for low dense quasi-trees, one expects to have a 
considerable number of support vertices. As a consequence, lower bounds
would be stronger, and during the search,  trial number of vertices can be
chosen to be close to lower bounds. Vice versa, for high-dense graphs, the dominance 
number is typically small. As a result, the search space is also small and
upper bounds are expected to be stronger. Then the trial values in the search
can be chosen to be close to upper bounds. In general, choosing an appropriate
starting point (the number of vertices) in the search can essential reduce
the search time. For an average middle-dense graphs, we choose $(L+3U)/4$ as the 
starting point in our search, since the existing lowers bounds are known
to be weak for such graphs.

%


\bibliography{mybibfile_enumerative}

\end{document}